\documentstyle[prl,aps,epsfig]{revtex}
\begin{document}

\draft \twocolumn[\hsize\textwidth\columnwidth\hsize\csname
@twocolumnfalse\endcsname

%\documentstyle[prd,eqsecnum,preprint,tighten,aps]{revtex}  % single space
%\documentstyle[prd,eqsecnum,preprint,aps]{revtex}           % double space for referee
%%%%\documentstyle[12pt]{article}
%\begin{document}                                          % OK for single and double space
%\renewcommand{\theequation}{\thesection.\arabic{equation}} %OK for single and double space
%%%\draft

 \title{Spin-Rotation Coupling in Muon $g-2$ Experiments}
\author{G. Papini$^{a,b}$\cite{papini@uregina.ca}}
\address{$^a$Department of Physics, University of Regina, Regina, Sask. S4S 0A2, Canada.}
\address{$^b$International Institute for Advanced Scientific Studies,
 84019 Vietri sul Mare (Sa), Italy.}
\author{G. Lambiase$^{c,d}$\cite{lambiase@sa.infn.it}.}
 \address{$^c$Dipartimento di Fisica "E.R. Caianiello",
 Universit\'a di Salerno, 84081 Baronissi (Sa), Italy.}
 \address{$^d$Istituto Nazionale di Fisica Nucleare, Gruppo Collegato di Salerno, Italy.}
\date{\today}
\maketitle
\begin{abstract}
Spin-rotation coupling, or Mashhoon effect, is a phenomenon associated with rotating
observers. We show that the effect exists and plays a fundamental role in the
determination of the anomalous magnetic moment of the muon.
\end{abstract}

\vskip2pc]

Fully covariant wave equations predict the existence of a new
class of inertial-gravitational effects that can be tested
experimentally. In these equations inertia and gravity appear as
external classical fields, but, by conforming to general
relativity, provide very valuable information on how Einstein's
views carry through in the world of the quantum. The absence of a
complete quantum theory of gravity and the difficulty of
acquiring experimental information at this level, render appealing
alternatives aimed at bridging the gap between general relativity
and quantum physics.

The unified treatment of inertia-gravity afforded by general
relativity ensures that there be complete correspondence between
gravitational and inertial effects whenever non-locality is not an
issue. For this reason inertia has at times played the harbinger
role in the study of gravitational effects.

Interest in inertia is also fostered by rapid experimental advances that require that
inertial effects be identified with great accuracy in precise Earth-bound and near space
tests of fundamental theories.

Experiments already confirm that inertia and Newtonian gravity
affect quantum particles in ways that are fully consistent with
general relativity down to distances of $10^{-3}$cm for
superconducting electrons \cite{hildebrandt,zimmerman} and
$10^{-13}$cm for neutrons \cite{colella,werner,bonse}.

Spin-inertia and spin-gravity interactions are the subject of
numerous theoretical
\cite{hehl,audretsh,huang,mashhoon,mashhoon1} and experimental
efforts \cite{ni,silverman,versuve,wineland,ritter}. Studies of
fully covariant wave equations carried out from different
viewpoints \cite{hehl-ni,singh,obukov,cai,cai91} identify
entirely similar inertial phenomena. Prominent among these is the
spin-rotation effect first described by Mashhoon \cite{mashhoon1}
who found that the Hamiltonians $H$ and $H'$ of a neutron in an
inertial frame $F_0$ and in a frame $F'$ rotating with angular
velocity $\omega$ relative to $F_0$, are related by
$\displaystyle{H'=H-\frac{\hbar}{2}\,{\vec \omega}\cdot {\vec
\sigma}}$.

This effect is conceptually important: it extends our knowledge of
inertial effects and it does so in an area of particle physics
that is quantum mechanical $par$ $excellence$. The relevance of
the effect to physical \cite{cai} and astrophysical
\cite{cai91,papini94} processes has already been pointed out.

No direct experimental verification of the Mashhoon effect has so
far been reported. The
purpose of our work is to show that the effect exists, is sizable
and plays an essential
role in precise measurements of the $g-2$ factor of the muon.

The experiment \cite{bailey,farley} involves muons in a storage ring consisting of a
vacuum tube, a few meters in diameter, in a uniform vertical magnetic field. Muons on
equilibrium orbits within a small fraction of the maximum momentum are almost completely
polarized with spin vectors pointing in the direction of motion. As the muons decay,
those electrons projected forward in the muon rest frame are detected around the ring.
Their angular distribution thence reflects the precession of the muon spin along the
cyclotron orbits.

Our thesis is best proven starting from the covariant Dirac
equation
 \[
 \left[i\gamma^\mu (x)(\partial_\mu+i\Gamma_\mu (x))-m\right]
 \psi(x)=0\,,
 \]
where $\Gamma_\mu (x)$ represents the spin connection and contains
the spin-rotation interaction. The Minkowski metric has signature
-2 and units $\hbar=c=1$ are used. The calculations are performed
in the rotating frame of the muon. Then the vierbein formalism
yields $\Gamma_i=0$ and
\begin{equation}\label{1}
  \Gamma_0=-\frac{1}{2}\,
  a_i\sigma^{0i}-\frac{1}{2}\,\omega_i\sigma^i\,,
\end{equation}
where $a_i$ and $\omega_i$ are the three-acceleration and three-rotation
of the observer
and
 \[
 \sigma^{0i}\equiv\frac{i}{2}\, [\gamma^0, \gamma^i]=i
 \left(\matrix{ \sigma^i & 0 \cr
                0 & -\sigma^i \cr }\right)\,
 \]
in the chiral representation of the usual Dirac matrices. The second term in (\ref{1})
represents the Mashhoon effect. The first term drops out. The remaining contributions to
the Dirac Hamiltonian, to first order in $a_i$ and $\omega_i$, add up to
\cite{hehl-ni,singh}
\begin{eqnarray}\label{hamiltonian-Ni}
  H &\approx & {\vec \alpha}\cdot {\vec p}+m\beta+\frac{1}{2}
  [({\vec a}\cdot {\vec x})({\vec p}\cdot {\vec \alpha})+
  ({\vec p}\cdot {\vec \alpha})({\vec a}\cdot {\vec x})] \\
  & & -{\vec \omega}\cdot \left({\vec L}+\frac{{\vec \sigma}}{2}\right)\,.
  \nonumber
\end{eqnarray}
For simplicity all quantities in $H$ are taken to be
time-independent. They are referred to a left-handed tern of axes
rotating  about the $x_2$-axis in the clockwise direction of
motion of the muons. The $x_3$-axis is tangent to the orbits and
in the direction of the muon momentum. The magnetic field is
$B_2=-B$. Only the Mashhoon term then couples the helicity states
of the muon. The remaining terms contribute to the overall energy
$E$ of the states, and we indicate by $H_0$ the corresponding
part of the Hamiltonian.

Before decay the muon states can be represented as
\begin{equation}\label{2}
  |\psi(t)>=a(t)|\psi_+>+b(t)|\psi_->\,,
\end{equation}
where $|\psi_+>$ and $|\psi_->$ are the right and left helicity
states of the Hamiltonian $H_0$ and satisfy the equation
 \[
 H_0|\psi_{+,-}>=E|\psi_{+,-}>\,.
 \]
The total effective Hamiltonian is $H_{eff}=H_0+H'$, where
\begin{equation}\label{3}
  H'=-\frac{1}{2}\,\omega_2\sigma^2+\mu B\sigma^2\,.
\end{equation}
$\displaystyle{\mu=\left(1+\frac{g-2}{2}\right)\mu_0}$ represents the total magnetic
moment of the muon and $\mu_0$ is the Bohr magneton. Electric fields used to stabilize
the orbits and stray radial electric fields can also affect the muon spin. Their effects
can however be cancelled by choosing an appropriate muon momentum \cite{farley} and will
not be considered.

The coefficients $a(t)$ and $b(t)$ in (\ref{2}) evolve in time according to
\begin{equation}\label{4}
  i\frac{\partial}{\partial t} \left(\matrix{ a(t) \cr
                b(t) \cr }\right)=M \left(\matrix{ a(t) \cr
                b(t) \cr }\right)\,,
\end{equation}
where $M$ is the matrix
\begin{equation}\label{5}
  M= \left[\matrix{ E-i\displaystyle{\frac{\Gamma}{2}} &
            \displaystyle{i\left(\frac{\omega_2}{2}-\mu B\right)}\cr
                \displaystyle{-i\left(\frac{\omega_2}{2}-\mu B\right)} &
                E-i\displaystyle{\frac{\Gamma}{2}} \cr }\right]
\end{equation}
and $\Gamma$ represents the width of the muon.

 $M$ has eigenvalues
\begin{eqnarray}
 h_1 &=& E-i\frac{\Gamma}{2}+\frac{\omega_2}{2}-\mu B \,, \nonumber \\
 h_2 &=& E-i\frac{\Gamma}{2}-\frac{\omega_2}{2}+\mu B \,, \nonumber
\end{eqnarray}
and eigenstates
\begin{eqnarray}
 |\psi_1> &=&
 \frac{1}{\sqrt{2}}\,\left[i|\psi_+>+|\psi_->\right]\,, \nonumber
 \\
 |\psi_2> &=& \frac{1}{\sqrt{2}}\,\left[-i|\psi_+>+|\psi_->\right]
 \,. \nonumber
\end{eqnarray}
The muon states that satisfy (\ref{4}) are
\begin{eqnarray} \label{6a}
 |\psi(t)> &=& \frac{e^{-\Gamma t/2}}{2}
 \left\{
 i\left[e^{-i{\tilde \omega} t}
 -e^{i{\tilde \omega} t}\right]|\psi_+> \right.
 \\
 & & \left. + \left[e^{-i{\tilde \omega} t}
 +e^{i{\tilde \omega} t}\right]|\psi_-> \right\}
 \,, \nonumber
\end{eqnarray}
where
 \[
 {\tilde \omega}\equiv E+\frac{\omega_2}{2}-\mu B\,.
 \]
At $t=0$ one has $|\psi (0)>=|\psi_->$. The spin-flip probability is therefore
\begin{eqnarray}\label{7}
  P_{\psi_-\to \psi_+}&=&|<\psi_+|\psi(t)>|^2 \\
     &=& \frac{e^{-\Gamma
  t}}{2}[1-\cos(2\mu B-\omega_2) t]\,. \nonumber
\end{eqnarray}
The $\Gamma$-term in (\ref{7}) accounts for the observed exponential decrease in electron
counts due to the loss of muons by radioactive decay \cite{farley}.

The spin-rotation contribution to $P_{\psi_-\to \psi_+}$ is
represented by $\omega_2$ which is the cyclotron angular velocity
$\displaystyle{\frac{eB}{m}}$ \cite{farley}. The spin-flip
angular frequency is then
 \begin{eqnarray}\label{omegafin}
 \Omega&=&2\mu B-\omega_2 \\
 &=&\left(1+\frac{g-2}{2}\right)\frac{eB}{m}-
 \frac{eB}{m} \nonumber \\
 &=& \frac{g-2}{2}\frac{eB}{m}\,, \nonumber
 \end{eqnarray}
which is precisely the observed modulation frequency of the
electron counts \cite{picasso} (see also Fig. 19 of Ref.
\cite{farley}). This result is independent of the value of the
anomalous magnetic moment of the particle. It is therefore the
Mashhoon effect that evidences the $g-2$ term in $\Omega$ by
exactly cancelling, in $2\mu B$, the much larger contribution
$\mu_0$ that pertains to fermions with no anomalous magnetic
moment.

It is perhaps odd that spin-rotation coupling as such has gone unnoticed for such a long
time. It is however significant that its effect is observed in an experiment that has
already provided crucial tests of quantum electrodynamics  and a test of Einstein's
time-dilation formula to better than a 0.1 percent accuracy. Recent versions of the
experiment \cite{carey,brown1,brown2} have improved the accuracy of the measurements from
270 ppm to 1.3 ppm. This bodes well for the detection of effects involving spin, inertia
and electromagnetic fields or inertial fields to higher order.

 This work was supported in part by the Natural Sciences and
Engineering Research Council of Canada.  G.L.'s research was supported by fund MURST PRIN
99.

%\newpage

\end{document}